# PARTICIPANT-INDEPENDENT VALIDATION OF SMARTPHONE VIDEO AND WEARABLE INERTIAL SENSING FOR COACH-RATED RUNNING-FORM CLASSIFICATION

**Kian Lee and Eric Zhou**
Northfield Mount Hermon School
Klee12260@gmail.com

## ABSTRACT

Automated running-form assessment could make technique feedback more accessible outside professional coaching and laboratory settings. We evaluated a low-cost pipeline combining side-view smartphone video with wearable inertial measurement unit (IMU) acceleration for coach-rated running-form classification. The dataset comprised 294 labeled recordings from 20 runners and 1,555 windows across four cues: landing position, suspension, triple extension, and upper-body posture. The primary analysis used 100 repeated four-runner holdout splits, recording-level aggregation of window predictions, and LightGBM models with hyperparameters fixed from prior modality-specific searches. Video-only LightGBM achieved mean macro F1 of 0.310, versus 0.168 for a majority-class baseline. Adding IMU features reduced macro F1 to 0.267, a paired mean difference of -0.043. Robustness analysis with Logistic Regression, Linear SVM, Random Forest, XGBoost, and LightGBM found that video-only models had higher mean performance than the majority baseline, whereas fusion effects depended on model family: Linear SVM showed a small positive effect, Logistic Regression was approximately neutral, and tree-based models showed lower performance with fusion. These results indicate a modest participant-independent signal in smartphone pose features. However, apparent IMU-fusion gains depended strongly on validation design, model family, and target category. The gap between recording-level and participant-independent performance underscores the need for runner-holdout evaluation and suggests that simple feature-level fusion does not reliably generalize to unseen runners.

## 1 INTRODUCTION

Running-related musculoskeletal injuries are common among recreational and competitive runners, but the biomechanical contributors to injury are multifactorial and are not captured by a single gait variable [1-3]. Prior work has shown that cadence, loading rate, vertical oscillation, joint loading, and step-rate manipulation can change mechanical variables linked to running injury risk [4,5,7]. Coaches and laboratory gait-analysis systems can identify visible patterns such as overstriding, reduced flight or suspension, insufficient lower-limb extension, and excessive trunk motion, yet instrumented gait analysis remains costly, time-intensive, and unavailable to many runners. Translating these observations into scalable feedback tools requires sensing systems that are affordable, portable, and valid outside controlled laboratory conditions.

Smartphone video and wearable sensors are plausible tools for this goal. Markerless pose methods, including OpenPose, DeepLabCut, MediaPipe, and BlazePose, have made body-landmark extraction from ordinary video more accessible [10-13]. Recent clinical and field-oriented work, including smartphone-based systems such as OpenCap, suggests that video-derived pose can support quantitative movement analysis, although markerless methods still require task-specific validation and careful interpretation [9,14,15]. Wearable inertial sensors have also been used to estimate running gait measures such as cadence, ground-contact time, tibial acceleration, shock attenuation, and segment kinematics in laboratory and field settings [16,19]. Machine-learning approaches have been explored for running-related injury prediction and risk classification from kinematic, kinetic, and wearable-sensor data [20,21]. Combining video and IMU data is an instance of multimodal feature-level fusion, which can improve modeling when modalities provide complementary signals but can also increase dimensionality, alignment error, and noise when standardization is weak [22]. Small biomechanical datasets are especially vulnerable to apparent performance gains that do not transfer to new participants, especially when runner identity, body shape, camera setup, footwear, and sensor placement can become shortcuts [23,24].

This study evaluates whether smartphone video and wearable IMU data can classify coach-rated running-form categories under participant-independent validation. The primary question is not whether the model is ready for deployment or whether it predicts future injury. Instead, the study asks whether observable running-form labels contain a learnable signal when evaluation is restricted to runners held out from training, and whether simple feature-level IMU fusion improves that participant-independent performance.

# 2 MATERIALS AND METHODS

## 2.1 PARTICIPANTS AND DATA COLLECTION

The full data collection consisted of 294 synchronized video and acceleration recordings from 20 runners. Participants ran at multiple speeds, approximately spanning 6 to 16 min/mi, to capture variation in running mechanics. The analyses reported here used the complete analyzable modeling subset with matched labels and windowed feature files.

Table 1. Class distribution in the analyzed modeling subset. Class values represent ordered coach-rated severity levels.

| **Outcome** | **Class 0** | **Class 1** | **Class 2** | **Total** |
|---|---|---|---|---|
| Landing position | 115 | 123 | 56 | 294 |
| Suspension | 64 | 57 | 173 | 294 |
| Triple extension | 55 | 138 | 101 | 294 |
| Upper-body posture | 65 | 124 | 105 | 294 |

## 2.2 SENSING HARDWARE AND VIDEO ACQUISITION

Each running trial was recorded from the sagittal view with a smartphone camera at 60 fps. The wearable subsystem used four MPU6050-based IMU nodes placed on lower-body and trunk-related segments, with devices on the shoe/foot, ankle/shin, lower thigh, and lower-back or pelvis region. The IMU nodes sampled at 60 Hz and transmitted data wirelessly through an nRF24L01-based receiver to a computer logger with timestamped rows. This configuration was intended to capture acceleration and timing information that wearable-sensor studies have linked to running gait analysis, while keeping the hardware low-cost and portable [16,19].

## 2.3 FEATURE CONSTRUCTION

Video frames were processed with MediaPipe Pose to extract body landmarks [12,13]. Pose features used normalized x-y landmark coordinates, with facial and selected extremity landmarks removed to emphasize trunk and lower-extremity mechanics. Because markerless pose estimates can vary with camera view, occlusion, clothing, lighting, and task context, the present analysis treated video-derived landmarks as candidate features rather than direct substitutes for marker-based laboratory kinematics [9,14]. To mitigate error caused by MediaPipe Pose, participants were asked to wear tight-fitting clothes with colors that contrasted the indoor background. IMU data were parsed into acceleration streams from the four sensor nodes. Video and IMU streams were aligned using first-frame timing and nearest-neighbor timestamp matching with gap filtering. The fused recordings were segmented into 45-frame windows, corresponding to 0.75 s at 60 fps, with a 10-frame stride. Video-only models used flattened pose features. Multimodal models concatenated pose features and IMU acceleration features at the window level.

## 2.4 COACH RATINGS AND TARGET CATEGORIES

Three professional coaches rated each video using ordered 0/1/2 severity labels. The average of the three ratings was used as the label. Four target categories were modeled: landing position, suspension, triple extension, and upper-body posture. These cues were selected because they represent observable sagittal-plane running-form constructs: foot placement at initial contact, stance and flight behavior, lower-limb extension, and trunk orientation. Related variables, including step rate, joint loading, vertical oscillation, and gait retraining targets, have been studied in running-injury and gait-feedback research [3-8]. Suspension was also included because contact timing, flight behavior, shock attenuation, and segment acceleration are variables that wearable inertial sensors can plausibly capture during running [16,19]. Landing position describes foot placement relative to the center of mass at initial ground contact. Suspension describes lower-limb flexion after contact and the runner's ability to absorb impact and control center-of-mass deceleration. Triple extension describes coordinated extension of the hip, knee, and ankle. Upper-body posture describes sagittal-plane trunk angle during stance and swing. The final machine-learning dataset used one label per recording per category.

## 2.5 MODEL TRAINING AND STATISTICAL ANALYSIS

The primary analysis used LightGBM classifiers [32] evaluated with participant-independent validation. One multiclass classifier was trained for each target category and modality. Hyperparameters were fixed from prior modality-specific grid-search summaries to avoid running a new grid search inside each repeated split. For each of 100 repeats, four runners were held out for testing and all remaining runners were used for training. Candidate splits were constrained so both training and test sets contained all three classes for each of the four targets. The held-out windows were aggregated to recording-level predictions by averaging class probabilities across windows from the same recording and selecting the class with the highest mean probability. Participant-level holdout validation was treated as the primary analysis because algorithm-validation studies warn that random splits, model selection, and correlated observations can lead to optimistic performance estimates when dependency structure is ignored [23-25].

Macro F1 was the primary metric because the targets were multiclass and imbalanced, and F-measure summaries are more informative than accuracy alone when class prevalence is uneven [26]. Mean macro F1 was computed across the four target categories for each repeated split. The primary paired comparison was the difference between video plus IMU and video-only macro F1 on the same held-out runners. Empirical 2.5th to 97.5th percentile intervals across the 100 repeats describe split-to-split variability. Bootstrap confidence intervals for mean paired deltas were computed by resampling repeated-split deltas [27]. A majority-class classifier, fit on the training runners for each split, was used as a baseline. Model metrics were computed with scikit-learn-compatible routines [28].

To test whether the LightGBM result was specific to one model family, the same 100 runner-holdout splits were repeated for Logistic Regression, Linear SVM [29], Random Forest [30], and XGBoost [31]. These models used fixed hyperparameters from the same prior modality-specific grid-search workflow used for the LightGBM analysis. Logistic Regression and Linear SVM used standardized features where the original tuned pipeline used scaling. Probabilistic models used mean window-level class probabilities for recording-level aggregation. Linear SVM does not provide native class probabilities, so recording-level predictions for that model were formed by averaging window-level decision scores before selecting the highest-scoring class.

A secondary recording-level split analysis was conducted for comparison with the participant-independent validation design. That split grouped windows by recording but did not hold out entire runners. It therefore estimates performance on unseen recordings rather than unseen participants and should be interpreted as a sensitivity analysis rather than the primary evidence for generalization [23,24].

# 3 RESULTS

## 3.1 DATASET SUMMARY

The analyzed subset contained 294 labeled recordings from 20 runners and 1,555 total windows. The median number of windows per recording was 5, with a range of 1 to 16. Class imbalance was present in all four target categories, with suspension showing the largest class-2 count.

## 3.2 PRIMARY PARTICIPANT-INDEPENDENT PERFORMANCE

Table 2. Primary repeated runner-holdout results. Intervals for baseline and model rows are empirical 2.5th to 97.5th percentile intervals across 100 splits; the paired-delta interval is a bootstrap confidence interval for the mean paired delta.

| Analysis | Mean macro F1 | Interval | Interpretation |
|---|---|---|---|
| Majority baseline | 0.168 | [0.114, 0.216] | Training-split majority class |
| Video-only LightGBM | 0.310 | [0.203, 0.470] | Pose features only |
| Video + IMU LightGBM | 0.267 | [0.153, 0.413] | Feature-level fusion |
| Paired delta | -0.043 | [-0.051, -0.035] | Bootstrap CI for mean delta |

Video-only LightGBM achieved mean macro F1 of 0.310, compared with 0.168 for the majority-class baseline. Video plus IMU achieved lower overall performance, with mean macro F1 of 0.267. The paired mean delta for video plus IMU relative to video-only features was -0.043, with bootstrap confidence interval [-0.051, -0.035]. Video plus IMU outperformed video-only in 11 of 100 repeated splits.

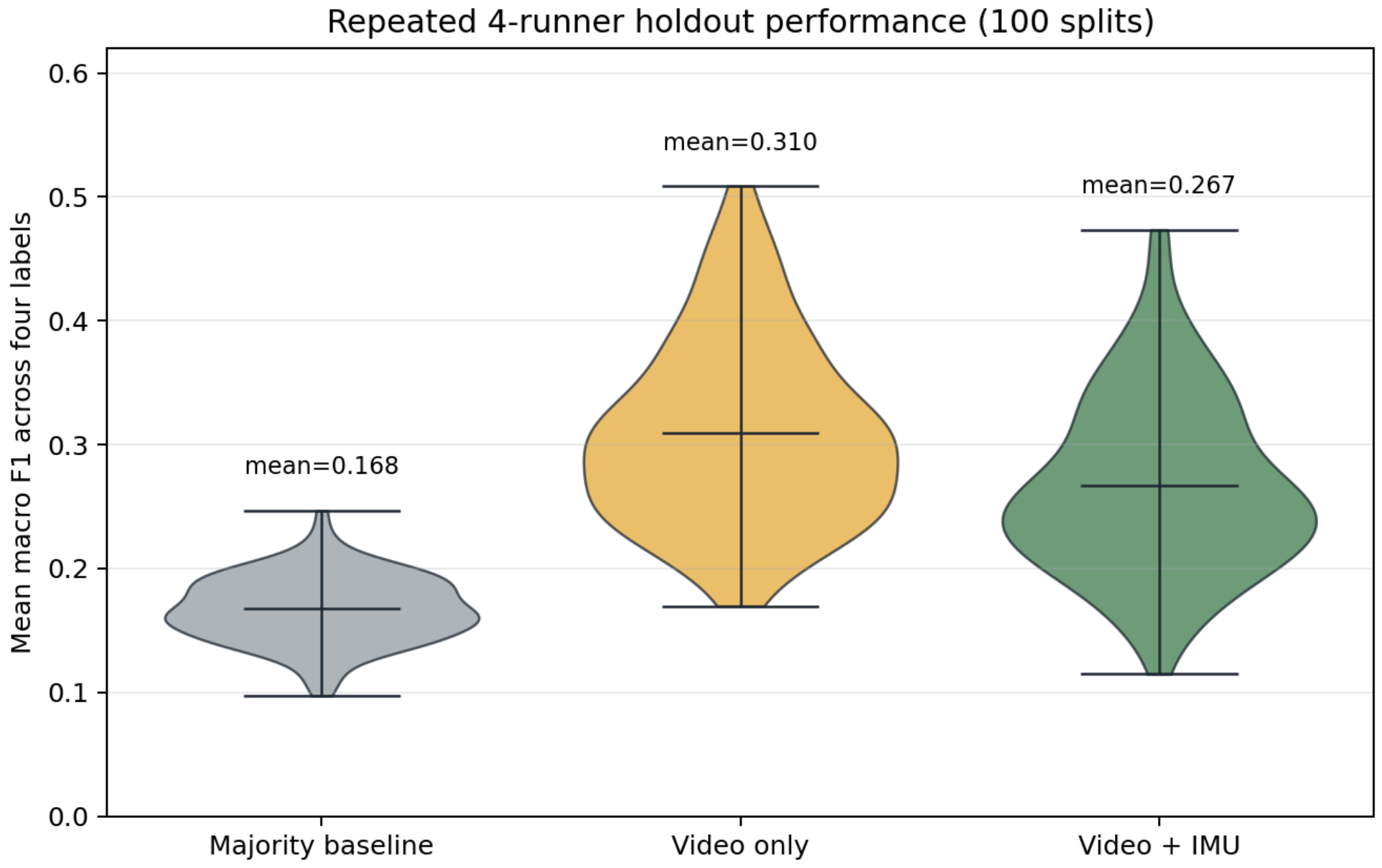

Figure 1. Distribution of mean macro F1 across 100 repeated four-runner holdout splits for the majority-class baseline, video-only LightGBM, and video-plus-IMU LightGBM.

### 3.3 CATEGORY-LEVEL FUSION EFFECTS

Table 3. Category-level repeated runner-holdout performance. F1 intervals are empirical 2.5th to 97.5th percentile intervals across splits; paired-delta intervals are bootstrap confidence intervals for mean video-plus-IMU minus video-only delta.

| Category | Video F1 | Video+IMU F1 | Paired delta | Positive Splits |
|---|---|---|---|---|
| Landing position | 0.378 [0.225, 0.569] | 0.348 [0.147, 0.624] | -0.031 [-0.047, -0.014] | 36/100 |
| Suspension | 0.313 [0.147, 0.503] | 0.355 [0.130, 0.619] | +0.041 [0.024, 0.060] | 64/100 |
| Triple extension | 0.281 [0.126, 0.503] | 0.226 [0.119, 0.379] | -0.056 [-0.071, -0.041] | 20/100 |
| Upper-body posture | 0.266 [0.095, 0.482] | 0.139 [0.024, 0.295] | -0.127 [-0.146, -0.109] | 6/100 |

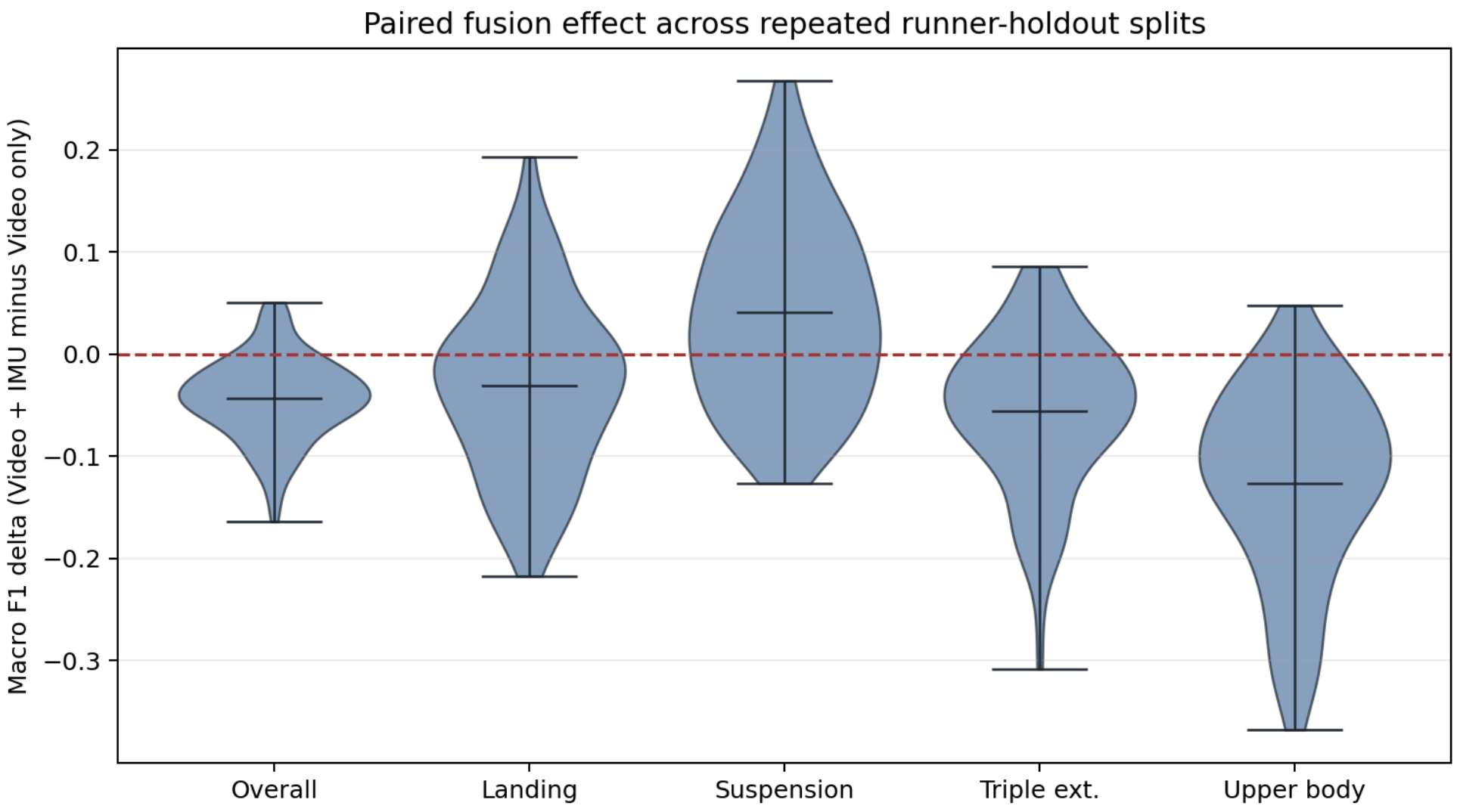

Figure 2. Paired macro F1 delta for video-plus-IMU minus video-only LightGBM across repeated runner-holdout splits. Values above zero indicate a fusion benefit.

The category-level analysis showed that the fusion result was not uniform. Video plus IMU improved suspension classification, but reduced performance for landing position, triple extension, and upper-body posture. The suspension result is biomechanically plausible because suspension depends on post-contact timing and acceleration patterns that may be more directly captured by IMU signals than by side-view pose alone.

### 3.4 ROBUSTNESS ACROSS MODEL FAMILIES

Table 4. Repeated runner-holdout model-family robustness analysis. The same 100 runner splits were used for every model. F1 intervals are empirical 2.5th to 97.5th percentile intervals across splits; paired-delta intervals are bootstrap confidence intervals for mean video-plus-IMU minus video-only delta.

| Model | Video F1 | Video+IMU F1 | Paired delta | Positive splits |
|---|---|---|---|---|
| Logistic Regression | 0.350 [0.222, 0.529] | 0.347 [0.239, 0.500] | -0.003 [-0.016, 0.009] | 51/100 |
| Linear SVM | 0.320 [0.200, 0.473] | 0.335 [0.228, 0.454] | +0.015 [0.003, 0.027] | 63/100 |
| Random Forest | 0.242 [0.154, 0.357] | 0.221 [0.133, 0.332] | -0.020 [-0.026, -0.015] | 26/100 |
| XGBoost | 0.287 [0.172, 0.412] | 0.257 [0.158, 0.391] | -0.030 [-0.037, -0.022] | 19/100 |
| LightGBM | 0.310 [0.203, 0.470] | 0.267 [0.153, 0.413] | -0.043 [-0.051, -0.035] | 11/100 |

The multi-model analysis supported the presence of participant-independent signal in video-derived pose features, but it did not support a uniform fusion benefit. Video-only mean macro F1 exceeded the majority baseline for every model family, ranging from 0.242 for Random Forest to 0.350 for Logistic Regression. In contrast, the fusion effect varied by classifier. Linear SVM showed a small positive overall fusion effect of +0.015, Logistic Regression was approximately neutral at -0.003, and Random Forest, XGBoost, and LightGBM all showed lower overall macro F1 with video-plus-IMU features.

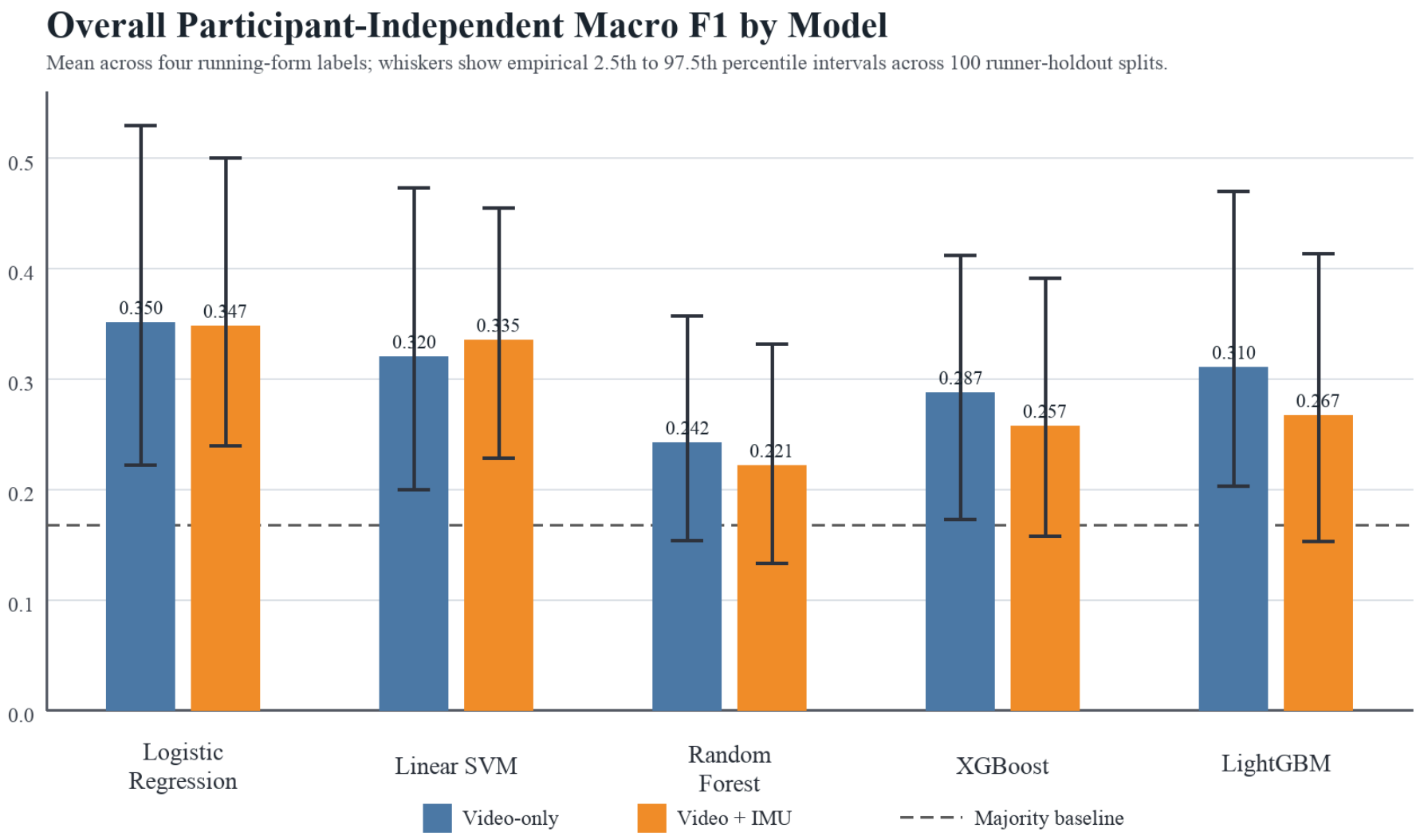

Figure 3. Overall mean macro F1 by model family and modality across the same 100 repeated four-runner holdout splits. Error bars show empirical 2.5th to 97.5th percentile intervals across splits.

Category-level deltas further showed that fusion effects were target-specific. Landing position improved with the linear models but decreased with XGBoost and LightGBM. Suspension showed the most consistent evidence of IMU benefit, with positive deltas for Linear SVM, XGBoost, and LightGBM. Triple extension and upper-body posture generally decreased with fusion, especially for the upper-body posture target. This pattern suggests that raw acceleration features may contain useful dynamic information for selected targets, while also introducing sensor-placement and participant-specific variability that harms other categories.

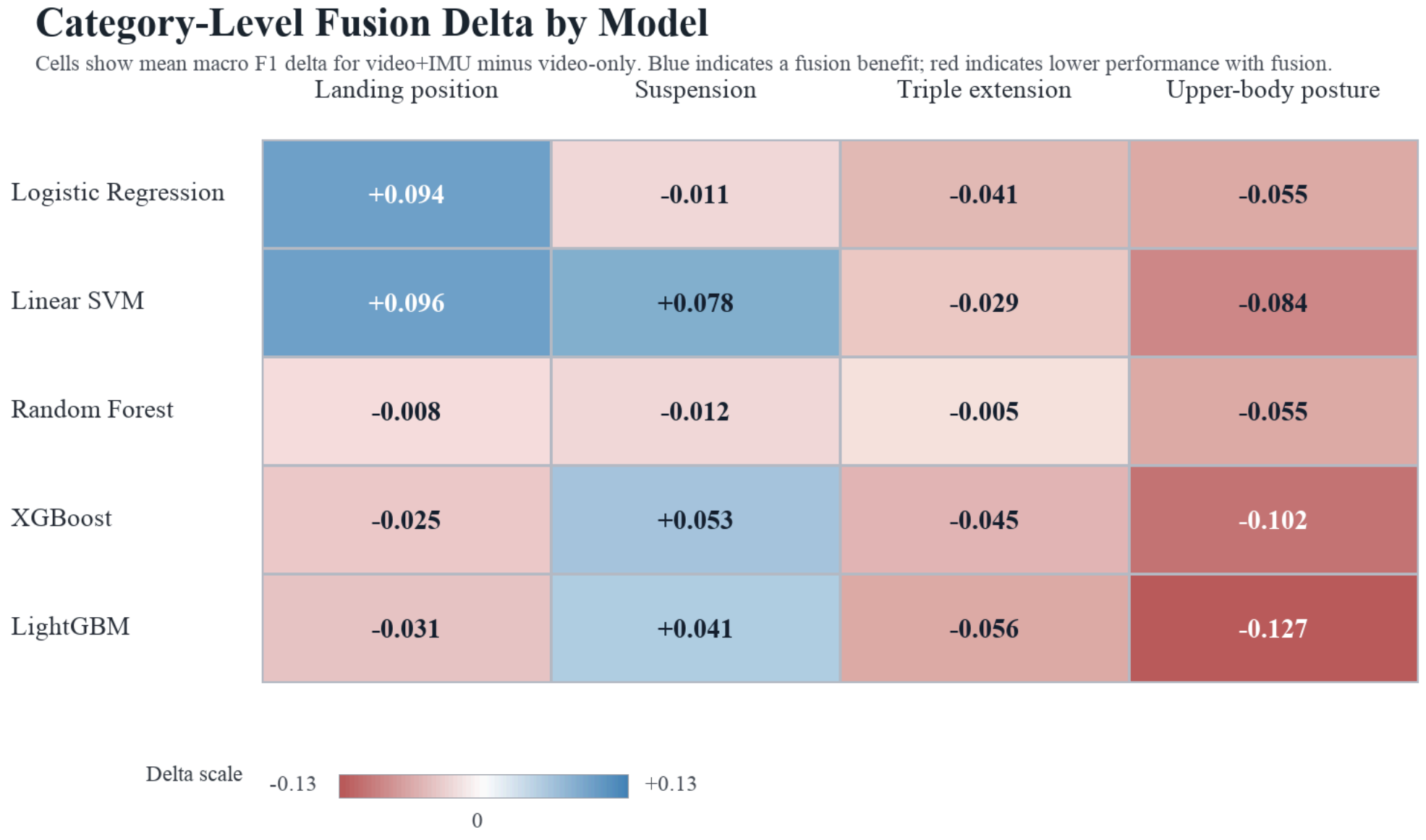

Figure 4. Category-level mean macro F1 delta for video-plus-IMU minus video-only features by model family. Blue cells indicate positive fusion effects and red cells indicate lower performance with fusion.

### 3.5 SECONDARY RECORDING-LEVEL SPLIT ANALYSIS

Table 5. Secondary comparison of participant-independent runner-holdout validation with the recording-level split. The recording-level split is retained as a sensitivity analysis because it can overestimate generalization when runners appear in both training and evaluation data.

| Validation design | Video F1 | Video+IMU F1 | Delta | Interpretation |
|---|---|---|---|---|
| Repeated runner-holdout | 0.310 | 0.267 | -0.043 | Tests unseen runners; primary analysis |
| Recording-level split | 0.640 | 0.734 | +0.094 | Tests unseen recordings, not unseen runners |

The recording-level split used 80% of the recordings for model training, and 20% for model evaluation. Under the recording-level split, video plus IMU LightGBM increased mean macro F1 from 0.640 to 0.734. This result contrasts with the participant-independent analysis, where video plus IMU reduced mean macro F1. The discrepancy indicates that recording-level validation can produce more optimistic fusion estimates when runner-level sources of variation are not held out.

## 4 DISCUSSION

This study found modest participant-independent signal in smartphone pose features, but it did not support a broad claim that feature-level IMU fusion improves running-form classification overall. Video-only models exceeded the majority-class baseline across all evaluated model families, indicating that the pose features captured some form-relevant information. However, mean macro F1 remained low across the analysis, so the current models should not be interpreted as reliable standalone classifiers.

The model-family robustness analysis indicated that the video-only signal was reproducible across model families, but the effect of IMU fusion was not. Linear SVM showed a small positive overall fusion effect, Logistic Regression was approximately neutral, and the tree-based models showed lower participant-independent performance after adding IMU features. Simple early fusion of raw acceleration and flattened pose features is not reliably beneficial under the present dataset, preprocessing, and validation design.

The most important methodological finding remains the gap between recording-level and runner-holdout validation. The recording-level split suggested that adding IMU features improved LightGBM performance. When entire runners were held out, overall fusion performance decreased for LightGBM and XGBoost and remained inconsistent across the broader model set. This pattern is consistent with runner-specific shortcuts or sensor-specific variation affecting the fused feature space, a known risk when correlated observations are split without respecting participant or group structure [23,24]. Small differences in IMU placement, orientation, attachment tightness, footwear, running style, and body morphology can change acceleration features in ways that may not transfer across participants.

The category-level results identify where IMU data may still be useful. Suspension was the clearest dynamic target, with positive fusion effects for Linear SVM, XGBoost, and LightGBM. Landing position improved for the linear models but decreased for boosted trees, suggesting that model capacity and feature scaling interact with the fused feature representation. Triple extension and upper-body posture generally decreased with fusion. The suspension result is biomechanically plausible because contact timing, flight time, impact attenuation, and acceleration patterns are among the types of signals wearable inertial sensors are designed to capture [16,19]. Wearable inertial sensors have also been studied for in-field sport-performance evaluation and indirect estimation of ground-reaction forces and moments [17,18]. Future work should therefore treat multimodal fusion as a modeling problem, not merely as concatenating more features.

These results should be interpreted as coach-rated running-form fault classification. The labels represent expert form ratings rather than prospective injury outcomes. Establishing injury-risk prediction would require longitudinal follow-up, injury definitions, training-load data, and prospective validation consistent with the broader running-injury literature [1-3]. The present analysis is therefore best viewed as a feasibility and validation study for automated running-form assessment.

### 4.1 LIMITATIONS

The modeling subset was modest in size, with 294 recordings from 20 runners, and should be expanded across age, sex, footwear, running surfaces, speeds, and training backgrounds.

Runner-holdout splits were constrained to contain all three classes for all four targets; this improves metric interpretability but does not replace a larger balanced dataset.

IMU axes and placement may not have been standardized enough for participant-independent fusion.

Side-view video emphasizes sagittal-plane mechanics and cannot capture all frontal- or transverse-plane motion.

Coach ratings are meaningful for training feedback but subjective; inter-rater reliability should be quantified in future work.

Classical model hyperparameters were fixed from prior modality-specific grid searches rather than re-tuned inside every repeated runner-holdout split.

Linear SVM does not provide native class probabilities, so its recording-level aggregation uses averaged decision scores rather than averaged probabilities.

## 5 CONCLUSION

These findings demonstrate that coach-rated running-form categories contain a modest participant-independent signal that can be extracted from smartphone pose features, as video-only models consistently outperformed the majority-class baseline across multiple model families. However, apparent gains from IMU fusion depended strongly on validation design, model family, and target category. The discrepancy between recording-level and participant-independent performance highlights the importance of runner-holdout evaluation and suggests that simple feature-level fusion does not reliably generalize to unseen runners. Suspension classification provided the clearest evidence that inertial signals may add target-specific value, but realizing that value will likely require improved sensor-placement standardization, preprocessing, and fusion strategies.

## DATA AVAILABILITY

The analysis code, fixed model hyperparameters, deidentified validation outputs, Python scripts, and figure-generating result tables are available in the public project repository on GitHub: https://github.com/kl1226/ml-rf-an-cb. Raw video recordings, raw inertial sensor logs, identifying filenames, local file paths, and the private participant-recording key are not released.

## ETHICS STATEMENT

This study involved human participants and was reviewed and approved before data collection by a school-level human-participant review committee convened in accordance with the Massachusetts Science & Engineering Fair and International Science and Engineering Fair human-participant research requirements. The study was also reviewed and approved by the Massachusetts Statewide Scientific Review Committee. All participants provided informed consent prior to participation, and parental or guardian consent was obtained for participants under the age of 18. Participants were informed that participation was voluntary and that they could stop at any time. Video and wearable-sensor data were stored securely, participant names were excluded from public reporting, and raw videos will not be shared because they may be identifiable.